\documentstyle[editedvolume,epsf]{crckapb}

\begin{opening}
\title{KINEMATICS OF NEARBY SUBDWARFS}

\author{B. FUCHS, H. JAHREISS, R. WIELEN}
\institute{Astronomisches Rechen-Institut\\
           M{\"o}nchhofstra{\ss}e 12-14, D-69120 Heidelberg, Germany}

\end{opening}

\runningtitle{NEARBY SUBDWARFS}

\begin{document}


\begin{abstract}
A sample of subdwarfs with accurate space velocities and standarized
metallicities is presented. This was constructed by combining Hipparcos
parallaxes and proper motions with radial velocities and metallicities
from Carney et al.~(1994; CLLA). The accurate Hipparcos parallaxes lead to
an - upward - correction factor of 11\% of the photometric distance scale
of CLLA. The kinematical behaviour of the subdwarfs is discussed in particular
in relation to their metallicities. Most of the stars turn out to be thick disk
stars, but the sample contains also many genuine halo stars. While the
extreme metal poor halo does not rotate, a population of subdwarfs
with metallicities in the range --1.6 $<$ [Fe/H] $<$ --1.0 dex appears to
rotate around the galactic center with a mean rotation speed of about 100 km/s.
\end{abstract}

\section{Data}

The data material, which we have analyzed, is based on the sample of high
proper motion stars by Carney et al.~(1994; CLLA). CLLA
have measured photometric parallaxes, radial velocities, and metallicities of
most of the A to early G stars, many late G and some early K stars in the {\it
Lowell Proper Motion Catalogue}. This data set of  1464 stars has been
cross-identified with the Hipparcos catalogue (ESA 1997) and we found 767
stars in common. However, the sample had to be cleaned under various
aspects. For instance, for some stars not all data were available or some
Hipparcos parallaxes were not accurate enough.
In Fig.~1 the colour-magnitude diagram of the remaining stars
is shown. Despite the efforts of CLLA to avoid them the sample is
still contaminated by giants and subgiants, which can be
detected now due to the improved accuracies of the absolute magnitudes. We have
removed all stars lying above a line in the CMD defined by the zero age main
sequence of stars with solar metallicity shifted upwards by $\Delta M_{\rm V}$
= 0.8 mag.
\begin{figure}[htbp]
\begin{center}
\epsfxsize=7cm
   \leavevmode
      \epsffile{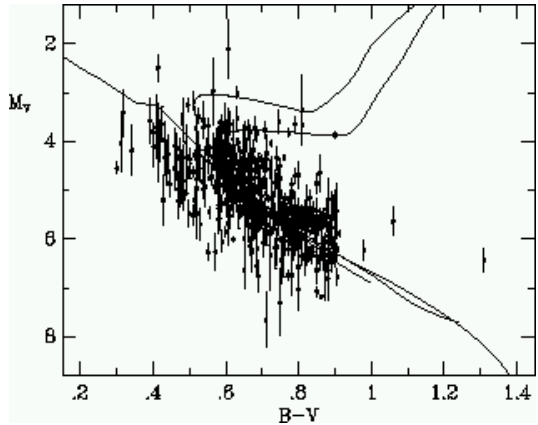}
\caption{CM-diagram of 617 identified CLLA stars. Hipparcos parallaxes were
used to determine the absolute magnitudes. The full lines indicate the
main sequence and the CMDs of the old clusters M\,67 and NGC\,188,
respectively.}    \label{fig1}
   \end{center}
   \end{figure}

Furthermore we have determined an overall correction factor of the photometric
distance scale of CLLA by analyzing parallax differences
$\pi_{\rm Hipp} - \pi_{\rm CLLA}$. A linear regression analysis
$\Delta\pi = f \cdot \pi_{\rm CLLA}$ gives a coefficient  $f$ = --0.113$\pm$
0.007, indicating
an upward correction of the photometric distances by 11\%. Jahrei{\ss} et
al.~(1997) have already found a similar correction on the basis of a smaller
sample of subdwarfs.

\section{Kinematics}

We have determined reliable space velocities of 560 nearby subdwarfs. These are
shown in Fig.~2 versus the metallicities of the stars as scatter plots.
The populations of thick disk and halo stars can be clearly distinguished by
the widths of the distributions and the mean rotational velocities. Since
our sample is based originally on a proper motion survey, the sample is biased
in the sense that low velocity stars are underrepresented. This can be seen
clearly as a trough in the U-velocity distribution in Fig.~2 or the asymmetric
V-velocity distribution of the disk stars in Fig.~3. However, since we are
mainly interested in the kinematics of the halo stars, this bias is of no
consequence in the present context.
\begin{figure}[htbp]
\begin{center}
\epsfxsize=12cm
   \leavevmode
      \epsffile{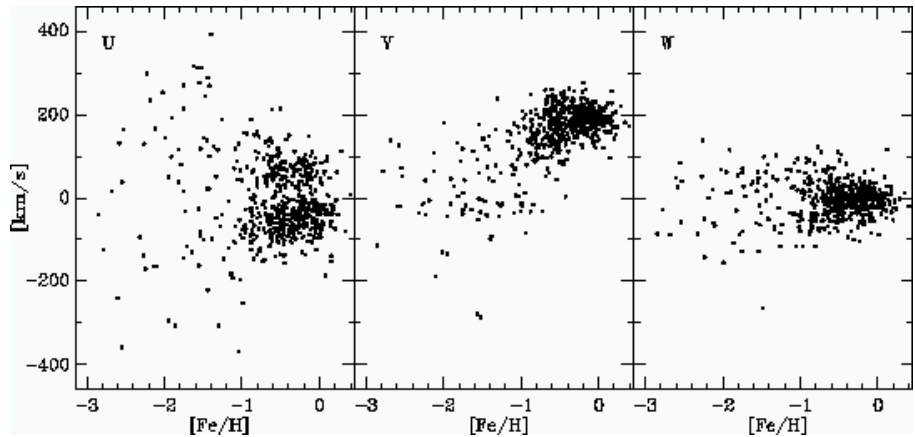}
\caption{Space velocity components versus metallicity [Fe/H]. U points towards
the galactic center, V in the direction of galactic rotation, and W towards the
galactic north pole. The space velocities have been reduced to the local
standard of rest and a rotation velocity of the LSR of 220 km/s has been
assumed.}
        \label{fig2}
   \end{center}
   \end{figure}

In order to determine mean velocities and velocity dispersions we have
considered three metallicity groups. Results are given in Table 1. The
corresponding V-velocity  distributions are
shown in Fig.~3. The first and largest group ([Fe/H] $>$ --1 dex)
represents obviously thick disk stars. The third group ([Fe/H] $<$ --1.6 dex)
has the same  kinematics as other tracers of the extreme
metal poor halo (cf.~the article by Norris in this volume). The second group
(--1.6 $<$ [Fe/H] $<$ --1)
shows peculiar kinematics. The V-velocity distribution is rather skewed with
respect to the mean. The V-velocity distribution at velocities less than
the mean velocity of the extreme metal poor halo of --220 km/s is
statistically indistinguishable
from that of the extreme metal poor halo, while at velocities larger than --220
km/s there is an excess of 15 stars compared to the corresponding distribution
of the extreme metal poor halo. Indeed, 19 out of the sample of 48 stars lie
beyond one standard deviation of the velocity distribution of the extreme metal
poor halo, i.e. at velocities less than --133 km/s.
We have carried out Monte-Carlo simulations, which show that such a distribution
has {\em not} been drawn with a probability of 99.99\% from
a gaussian distribution resembling that of the extreme metal poor halo. This
is illustrated further in the fourth panel of Fig.~3, where we show the
resulting distribution, if the left part of the V-velocity distribution is
folded at V = --220 km/s, subtracted from the right part of the distribution
and smoothed by a running mean.
This gives the distinct impression of an excess
population of subdwarfs in this metallicity range, which rotates with a
velocity of about 100 km/s around the galactic center, while the
rest of the subdwarfs in this metallicity range have the same kinematics
as the extreme metal poor halo.
\begin{table}[htbp]
\begin{center}
\caption{Kinematic parameters}
\begin{tabular}{cccccccc}
\hline
   & N & $\overline{U}$ & $\overline{V}$ & $\overline{W}$ & $\sigma_{\rm U}$
& $\sigma_{\rm V}$ & $\sigma_{\rm W}$\\
 dex & & \multicolumn{6}{c}{km/s} \\
\hline
{[Fe/H]}$>$ --1           & 477 & --21 & --54 & --10  & 72  & 46  & 38 \\
--1.6 $<$ {[Fe/H]} $<$ --1 & 48  & 6   & --186 & --17 & 168 & 105 & 73 \\
{[Fe/H]} $<$ --1.6         & 35  & 4   & --216 & --15 & 181 & 83 & 74 \\
\hline
\end{tabular}
\end{center}
\end{table}
\begin{figure}[htbp]
\begin{center}
\epsfxsize=6cm
   \leavevmode
      \epsffile{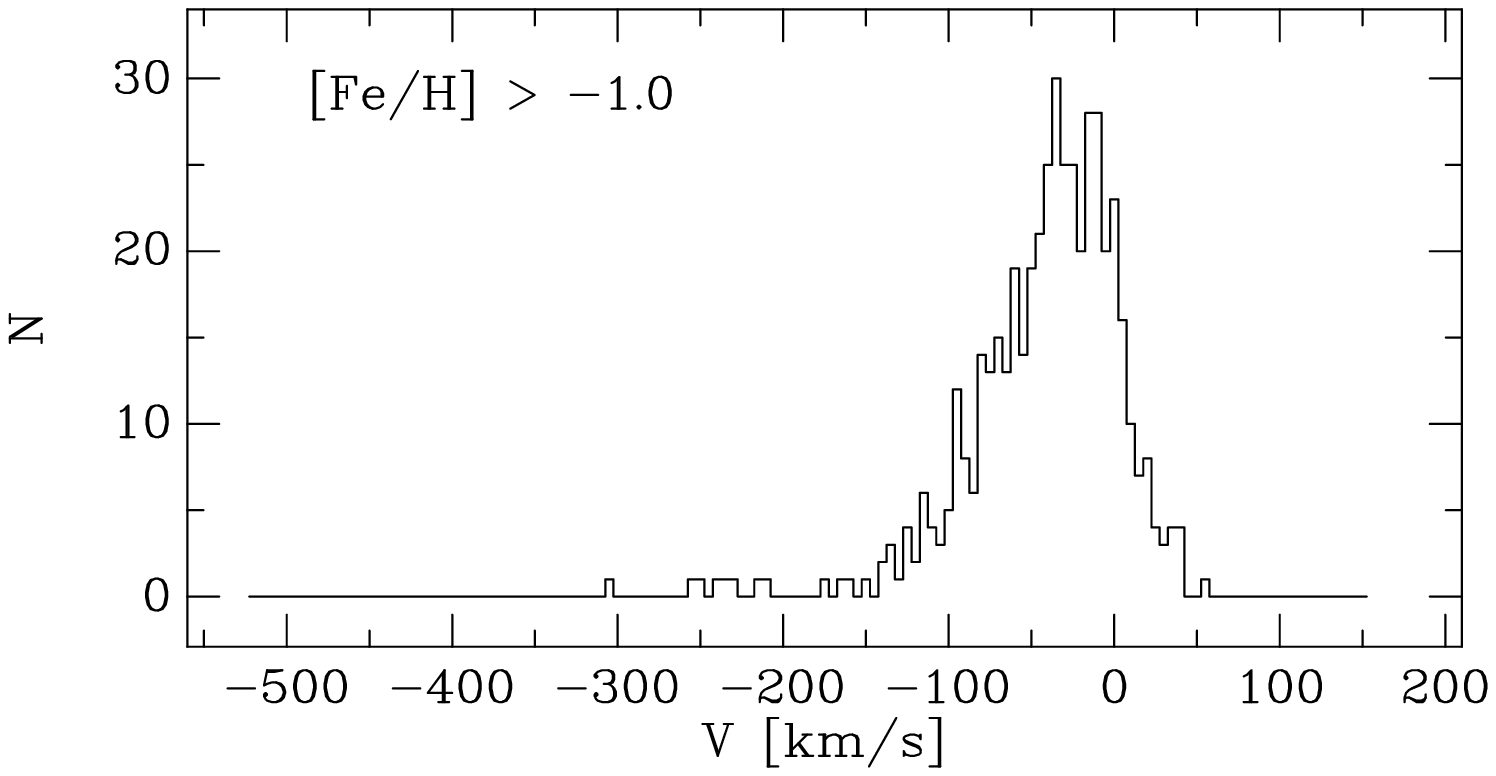}
      \epsfxsize=6cm
   \leavevmode
      \epsffile{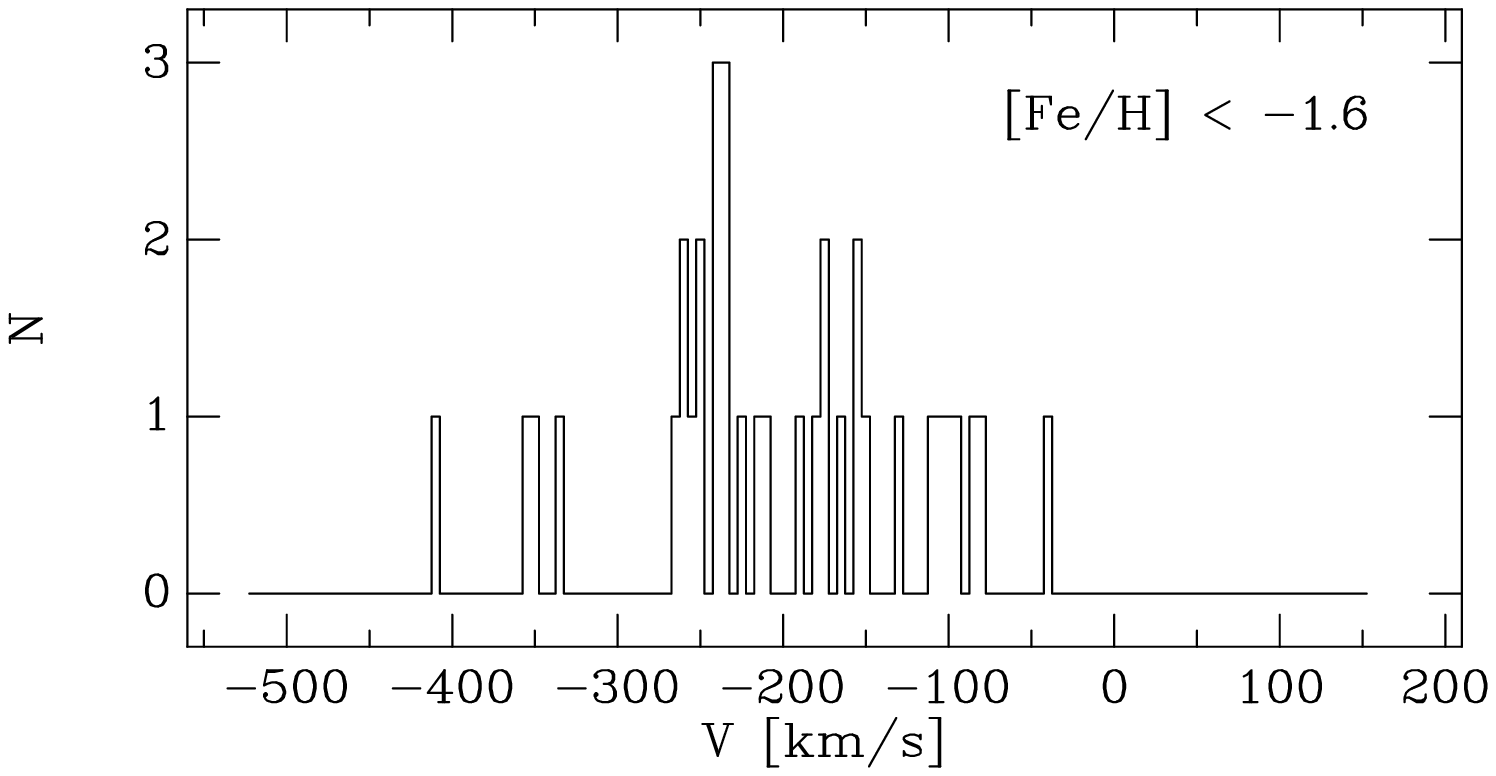}\\
\epsfxsize=6cm
   \leavevmode
      \epsffile{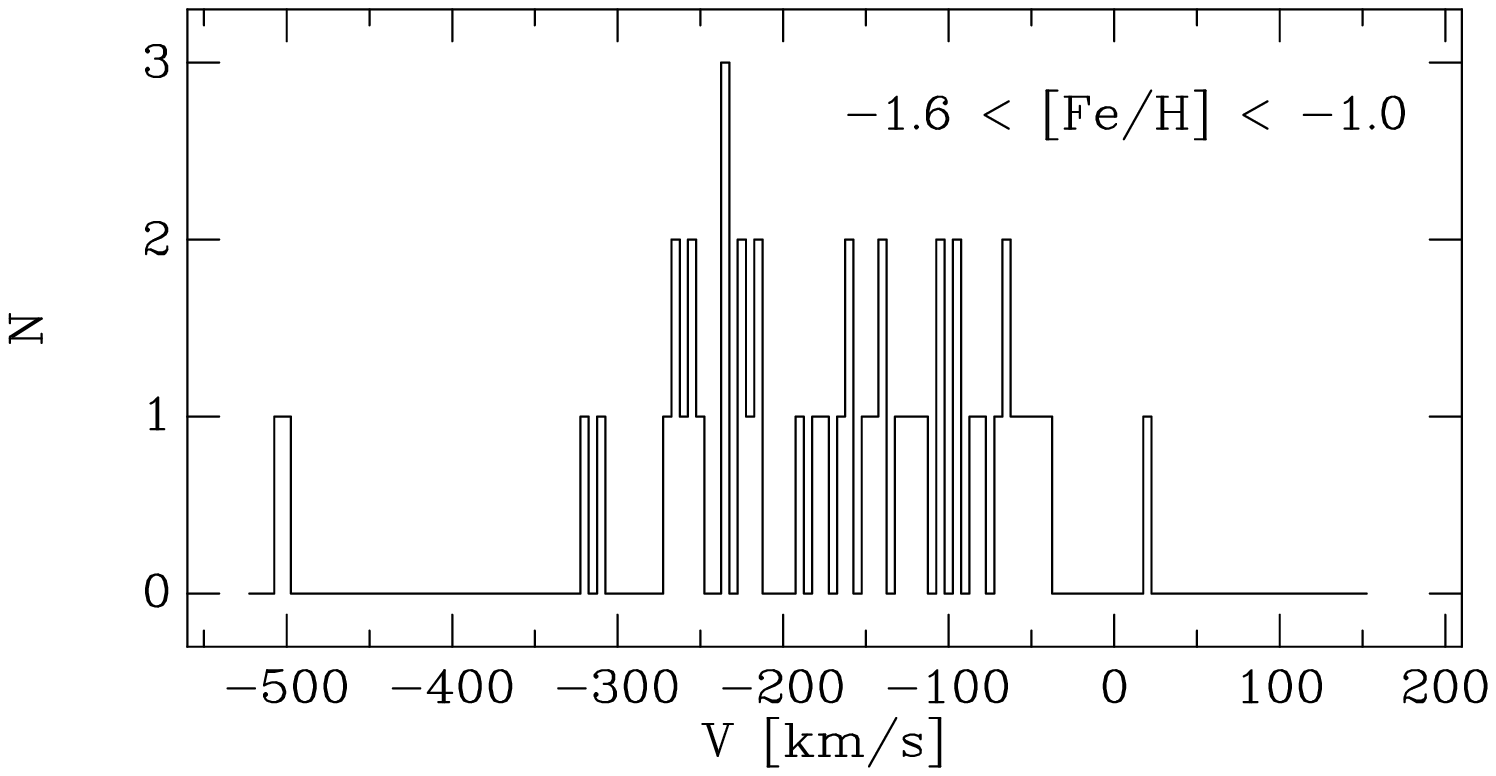}
      \epsfxsize=6cm
   \leavevmode
      \epsffile{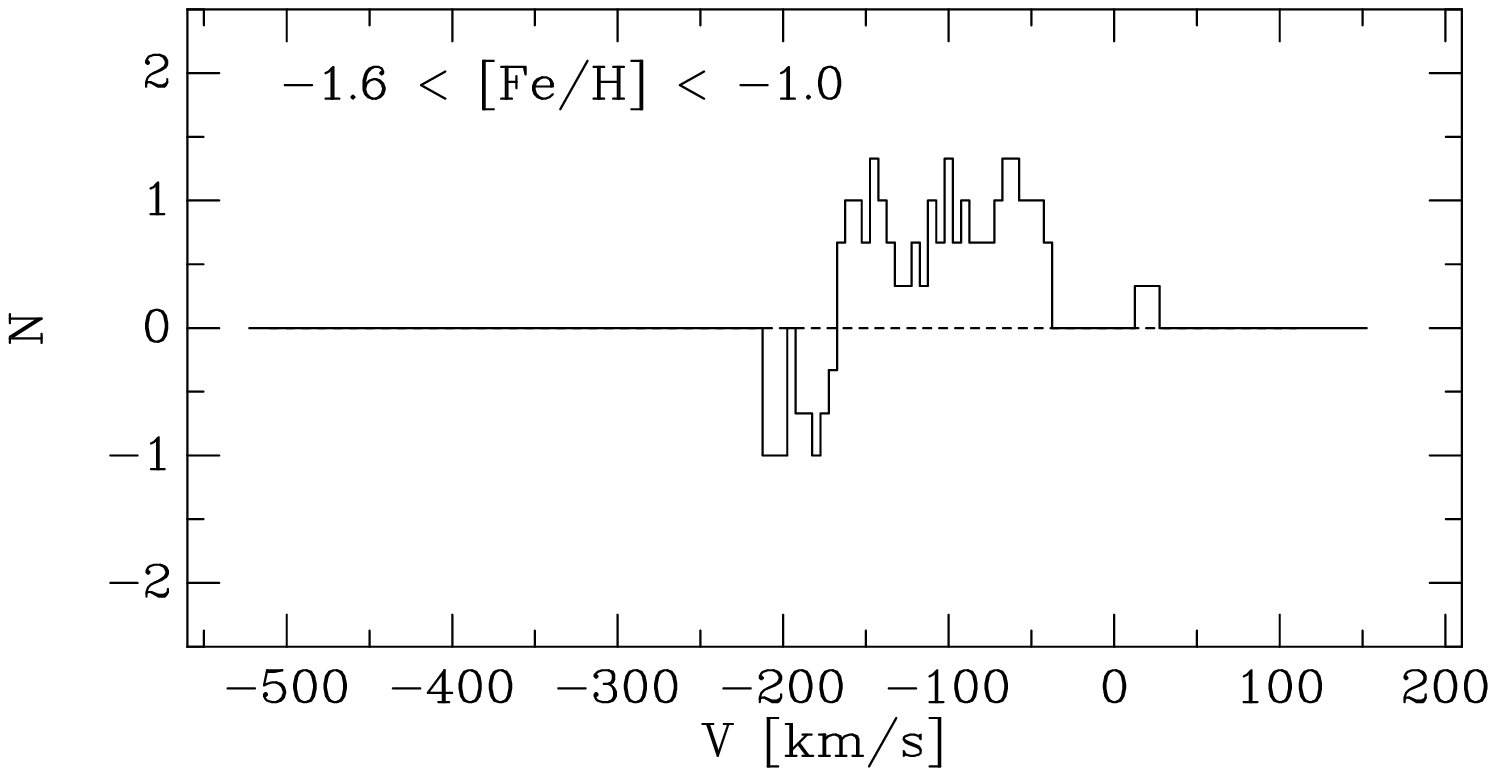}
\caption{V-velocity distributions of the subdwarfs grouped according to
their metallicities. The velocities are reduced to the local standard of rest.
The panel in the lower right shows the excess population (see text for
details).}
        \label{fig3}
   \end{center}
   \end{figure}

The nature of this excess population is not clear at present. It may well be
that these stars represent a metal weak --  and dynamically hot -- tail of the
thick disk. It is conspicuous that their asymmetric drift coefficient,
$\overline{V}/\sigma_{\rm U}^2$ = --1.2$\cdot$10$^{-2}$ (km/s)$^{-1}$, is very
similar to that of the thick disk $\overline{V}/\sigma_{\rm U}^2$ =
--1$\cdot$10$^{-2}$ (km/s)$^{-1}$. According to this interpretation late stages
of halo formation and early phases of disk formation would have overlapped.
Alternatively, this population may be a relic of earlier
accretion events (cf.~the article by Majewski in this volume).

\end{document}